\documentclass[reprint,aps,pra,showpacs,preprintnumbers, showkeys,amsmath,amssymb,floatfix,superscriptaddress]{revtex4-2}
\usepackage{graphicx}
\usepackage{color,colortbl}
\usepackage{xcolor}
\usepackage[normalem]{ulem}
\usepackage{amsfonts}
\usepackage[breaklinks=true,colorlinks=true,linkcolor=blue,urlcolor=blue,citecolor=blue]{hyperref}
\usepackage{etoolbox}
\usepackage{physics}
\usepackage{silence}

\apptocmd{\thebibliography}{\raggedright}{}{}
\makeatletter 
\def\@bibdataout@aps{%
\immediate\write\@bibdataout{%
  @CONTROL{%
  apsrev41Control,author="48",editor="1",pages="0",title="",year="1"%
  }%
}%
\if@filesw
  \immediate\write\@auxout{\string\citation{apsrev41Control}}%
\fi
}%
\makeatother

\usepackage{tikz,xcolor,hyperref}

\definecolor{lime}{HTML}{A6CE39}
\DeclareRobustCommand{\orcidicon}{%
	\begin{tikzpicture}
	\draw[lime, fill=lime] (0,0) 
	circle [radius=0.16] 
	node[white] {{\fontfamily{qag}\selectfont \tiny ID}};
	\draw[white, fill=white] (-0.0625,0.095) 
	circle [radius=0.007];
	\end{tikzpicture}
	\hspace{-2mm}
}

\foreach \x in {A, ..., Z}{%
	\expandafter\xdef\csname orcid\x\endcsname{\noexpand\href{https://orcid.org/\csname orcidauthor\x\endcsname}{\noexpand\orcidicon}}
}


\begin{document}

\title{Single attosecond XUV pulse source via light-wave controlled relativistic laser-plasma interaction: Thomson Back Scattering Scheme}

\author{Mojtaba Shirozhan}
\affiliation{ELI ALPS, ELI-HU Non-Profit Ltd., Wolfgang Sandner utca 3., Szeged H-6728, Hungary}
\affiliation{Institute of Physics, University of Szeged, D\'om t\'er 9, H-6720 Szeged, Hungary}
\author{Fabien Qu\'{e}r\'{e}\orcidB{}}
\affiliation{ELI ALPS, ELI-HU Non-Profit Ltd., Wolfgang Sandner utca 3., Szeged H-6728, Hungary}
\affiliation{Pasqal, 24 avenue Emile Boudot, 91120 Palaiseau, France}
\author{Subhendu Kahaly\orcidA{}}
\email[Corresponding author:]{\,subhendu.kahaly@eli-alps.hu}
\affiliation{ELI ALPS, ELI-HU Non-Profit Ltd., Wolfgang Sandner utca 3., Szeged H-6728, Hungary}
\affiliation{Institute of Physics, University of Szeged, D\'om t\'er 9, H-6720 Szeged, Hungary}

\date{\today}

\begin{abstract}
Reflecting light off a mirror moving near light speed offers a powerful method for generating bright, ultrashort pulses in the extreme ultraviolet range. Several investigations show that dense relativistic electron mirrors can be created by striking a nanometre-scale foil with a high-intensity, sharp-front laser pulse, forming a single relativistic electron sheet (RES). This RES coherently reflects and upshifts a counter-propagating laser beam from the infrared to the extreme ultraviolet with efficiency exceeding incoherent scattering by over several orders of magnitude. Here we demonstrate that optimizing the drive laser waveform can reliably produce a single RES, leading to generation of isolated \emph{attosecond} pulses enhancing both intensity and temporal compression of the back reflected light in a controlled manner. Simulations reveal that tuning parameters like timing delay enables control over the amplitude, duration, and bandwidth of the resulting attosecond Thomson backscattering pulse. Together, these advances meet key experimental challenges and pave the way for compact, tunable sources of isolated attosecond pulses for probing ultrafast phenomena.
\end{abstract}
\maketitle

\section{Introduction}\label{sec:intro}

Recent breakthroughs in ultrafast laser technology have enabled the generation of high-peak-power, few-cycle laser pulses spanning a broad spectral range \cite{Veisz2025}, enabling access to waveform-controlled relativistic laser plasma interaction. The availability of such extreme transient light fields together with advanced interaction platforms \cite{Mondal2018,Shirozhan2024} opens up the possibility of utilizing advanced secondary sources based on tunable quasimonoenergetic ion acceleration \cite{DeMarco2023} and intense attosecond (\emph{as}) pulses in the extreme ultraviolet (XUV) spectral range with controlled polarization \cite{Chen2016} or angular momentum \cite{Wang2019}, in real experiments.

Laser-plasma-based ultrashort XUV pulse sources, relying on high harmonic generation (HHG), have spurred significant research efforts due to their potential for probing matter with high temporal and spatial resolution. HHG resulting from the relativistic interaction of intense laser pulses with solid density plasma mirror targets has been shown to produce XUV beams with high conversion efficiency \cite{Yeung2016,Jahn2019,Kim2023} and a higher intensity of the drive laser yields a brighter XUV pulse, consisting of higher photon energies (see \cite{Edwards2020,Nayak2019} and the references therein). Driven by the multi-cycle intense laser pulse, the Doppler up-shifted frequency radiation can be potentially produced either in the specular reflection off the plasma surface \cite{Vincenti2014,Chopineau2021} or in the laser transmission direction \cite{Dromey2012}, coherently emitted by oscillating nano-meter scale electron bunches, which undergo a curved trajectory \cite{Mikhailova2012,Cousens2020}. Due to the periodicity of the generation process, this in general leads to \emph{as} pulse train in the time domain.  Isolated pulses \emph{as} are typically achieved by gating the generation either temporally or spatially to confine the interaction to only one intense optical cycle \cite{Rykovanov2008,Wheeler2012,Yeung2013}. In this process, the driving laser pulse self-consistently generates and interacts with relativistic electron bunches, leading to the emission of attosecond pulses.

An alternative and more controllable approach involves decoupling the generation of relativistic electron bunches from the attosecond pulse generation process. This separation allows for (i) independent generation and optimization of the electron bunch quality and (ii) a separate backscattering interaction for flexible tuning of the resulting XUV attosecond emission. Thomson backscattering of high-intensity laser pulses by relativistic electron bunches—long recognized as a robust mechanism for x-ray generation \cite{Lee2003, Li2015, Yan2017}—is founded on this principle. In this context, coherent Thomson backscattering (CTS) \cite{Kulagin2004, Kulagin2007} offers a compelling route to simultaneously upshift the carrier frequency and compress an entire femtosecond (fs) laser pulse into a single attosecond burst with relativistic intensity. This method inherently eliminates the need for temporal gating or spectral filtering, as the full pulse is coherently reflected and compressed. To ensure coherence, however, the electron bunch must be ultrathin—on the order of the emitted wavelength, typically a few nanometers. When such a dense electron sheet is accelerated to velocities approaching the speed of light, it effectively acts as a relativistic mirror.

To generate isolated \emph{as} pulses via CTS, the interaction must be confined to a single reflection event from a relativistically moving electron bunch \cite{Wu2009_443}. This requires the formation of a solitary, ultra-thin, and coherent electron mirror propagating at relativistic velocities, with which a counter-propagating laser pulse can interact. A well-established method for producing high-quality relativistic electron bunches is laser wakefield acceleration (LWFA), where intense, ultra-short laser pulses interact with underdense plasma. In this regime, electrons are accelerated to relativistic energies and can subsequently interact with a counter-propagating laser pulse \cite{Bulanov2003, Schwoerer2006, Pirozhkov2007,TaPhuoc2012}. Although LWFA can produce highly energetic electron bunches, achieving the high peak electron densities required for the formation of a RES remains a significant challenge. One promising approach, not based on LWFA, involves the interaction of intense laser pulses with nanometer-scale solid-density foil targets. This method can generate relativistic electron bunches with optimized density and energy distributions, enabling the production of intense attosecond pulses. In such a backscattering scheme, two counter-propagating laser pulses interact with an ultra-thin solid foil \cite{Kulagin2004, Kulagin2007}. The drive pulse must be sufficiently intense to expel electrons from the foil and accelerate them to relativistic velocities before the target disintegrates. The resulting overdense, nanometer-scale RES, when illuminated by the counter-propagating few-cycle source pulse, emits spatially coherent, Doppler-upshifted radiation with sub-femtosecond temporal duration. Although, the use of multiple RESs for high harmonic generation (HHG) has been demonstrated experimentally with multi-cycle, linearly polarized laser pulses \cite{Kiefer2013}, the generation of a single dense RES capable of producing isolated attosecond pulses remains an open challenge.

In this work, we investigate how the waveform of a linearly polarized drive laser influences the formation and properties of relativistic electron sheets. Through detailed analysis, we illustrate that an ideally sharp-fronted laser pulse can efficiently accelerate surface electrons into a single, dense, and coherent bunch. In contrast, a conventional pulse with Gaussian temporal profile — commonly used in experiments—induces surface electron oscillations rather than collective acceleration, thereby failing to generate a well-defined RES. We overcome this limitation, by tailoring the drive laser waveform via introducing a controlled second harmonic component to the fundamental Gaussian pulse. This two-color configuration, implementable in experiments, enables precise shaping of the electric field, resulting in an optimized waveform capable of producing a single, high-density RES. We demonstrate that such a relativistic electron mirror can then be employed to reflect a counter-propagating few-cycle pulse via the CTS mechanism, leading to the generation of intense, isolated attosecond pulses with controlled properties. Using particle-in-cell (PIC) simulations, we systematically explore the influence of various drive pulse waveforms and temporal envelopes on the interaction with ultrathin foil target, and its implication on the characteristics of the resulting backscattered radiation. Finally, we identify and present the optimal conditions under which the scattered pulse achieves maximal temporal compression and spectral upshift, paving the way for efficient generation of isolated attosecond bursts.

\section{Influence of drive laser waveform on TBS condition in laser-foil interaction}\label{sec:intro_TBS}

The generation of isolated attosecond pulses via Thomson basckscattering (TBS) from RES requires several stringent conditions. Foremost among these is the need for the RES to act as a coherent, mirror-like reflector. This necessitates that the electron sheet be significantly thinner than the wavelength of the reflected radiation in its rest frame, enabling phase-synchronized scattering and thus coherent emission. The efficiency of TBS in producing isolated attosecond pulses under these conditions is critically governed by the spatial and energy distribution of electrons within the RES, as well as by the temporal and spectral characteristics of the incident ultrashort laser pulse.

To enable the formation of an optimal Relativistic Electron Sheet (RES), the incident drive laser must exert a sufficiently strong ponderomotive force to expel plasma electrons from the ion background and accelerate them coherently in the forward direction. This requires the laser to generate a longitudinal push force capable of compressing electrons into a sub-skin-depth layer, ultimately overcoming the electrostatic restoring force of the ions. The necessary condition for this "snowploughing" \cite{Vincenti2014,Iwata2018} effect to enable complete electron evacuation, is that the normalized vector potential of the drive laser satisfies, $a_{0}^{d} > \omega_{p}^{2}d/c\omega_{L}$, where $\omega_p=\sqrt{n_{e}e^{2}/\epsilon_{0}m_{e}}$ is the plasma frequency, $d$ is the initial thickness of the plasma foil, $\omega_{L}$ is the drive laser frequency and $c$ is the speed of light  \cite{Vshivkov1998, Qiao2009, DeMarco2023, Kiefer2013}.

This electron blowout mechanism can be achieved either through a smoothly rising, few- or multi-cycle circularly polarized intense laser pulse (\cite{DeMarco2023} and references therein), which drives all electrons into a coherent sheet, or via a sharply rising (non-adiabatic) linearly polarized few-cycle driver \cite{Kulagin2007}, which confines the RES formation to sub-cycle durations. Once the electron sheet is formed, the drive pulse continues to propagate through the plasma, imparting a nearly uniform ponderomotive force across the ultra-thin, dense electron layer. This results in synchronous forward acceleration of all electrons within the sheet, preserving coherence and enabling efficient RES dynamics.

\begin{figure}[!tbh]
    \centering
    \includegraphics[width=1\columnwidth]{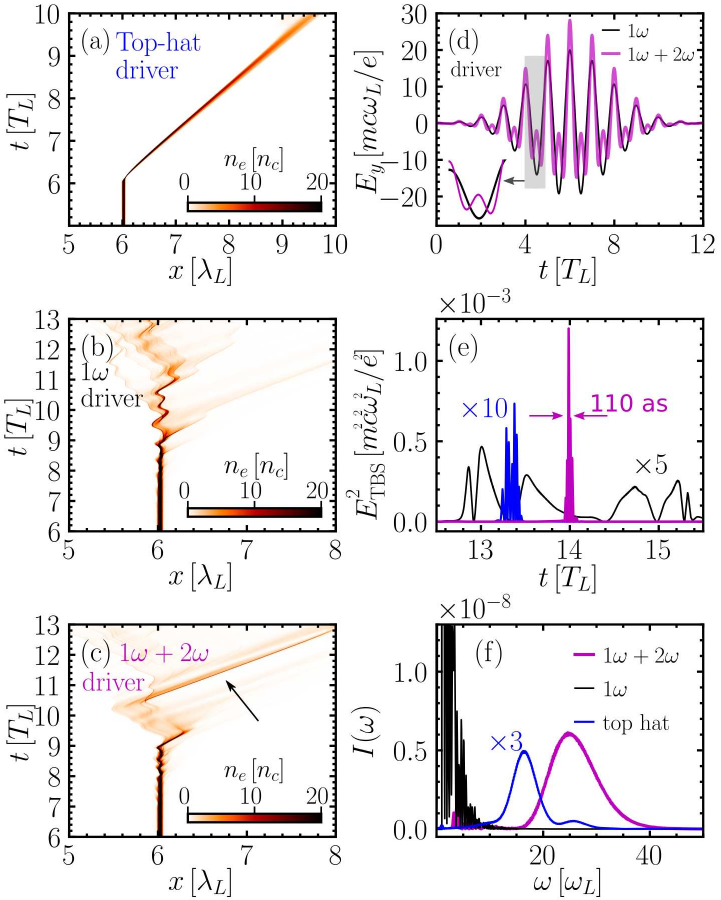}
    \caption{Influence of drive pulse temporal envelope and its waveform on the formation of the electron mirror and backscattered signal. The $40$nm thick plasma slab (located from $6$ to $6.05\lambda_{L}$) is irradiated by the drive pulse possessing different waveform and temporal envelope (a) top-hat, (b) Gaussian, and (c) optimal dual-color Gaussian (with $P=0.4$) pulse, which causes various temporal evolution of electron number density. In all cases the drive beam propagates from left to right and is incident normally on the $40$nm thick pre-ionized plasma foil at $6T_L$. (d) compares the temporal waveforms of the single ($1\omega$) and two-color ($1\omega+2\omega$) drive Gaussian pulses with end-to-end $12T_L$ temporal duration and polarization along y-direction. The inset at the lower left-hand side enlarges the waveforms from $4$ to $5T_{L}$, highlighting the steepness of accelerating two-color field. The counter propagating source pulse with duration $6T_L$ and peak amplitude of $a_0=0.1$ interacts with the electron sheet and reflects partially. The squared field of the normalized TBS signals are shown in (e), indicating the generation of a single $110$as pulse (in FWHM) in case of $(1\omega + 2\omega)$ drive beam. The duration of the TBS signal, resulting from interaction with the produced RES in panel (a) is $380$as. This pulse is scaled by a factor of $10$ (in blue). The reflected pulses are temporally shifted for better comparison. The corresponding frequency spectra of the TBS radiations are illustrated in panel (f). The AP generated by interaction of dual-color drive beam and foil target is composed of spectrum peaked at $26\omega_L$.}
    \label{fig:gau_2w}
\end{figure}

To gain a comprehensive understanding of how the temporal waveform of the drive laser pulse influences the interaction dynamics with a dense plasma foil, we carry out PIC simulations \cite{Lichters1996}. We examine the spatio-temporal evolution of the electron number density $n_{e}(x,t)$ (normalized by critical density $n_c$) as shown in Figure ~\ref{fig:gau_2w}(a). This figure depicts the interaction between a top-hat laser pulse—represented as a sinusoidal wave—and a nanometric ($d=40~nm$) solid-density foil. The incident laser is linearly polarized (electric field along the y-axis) and propagates along the x-axis (left to right) with normalized electric field amplitude of $a_0^{d}=eE_0^{d}/mc\omega_{L}=20$ corresponding to a peak intensity of $8.6 \times 10^{20}$W/cm$^{2}$ for a central wave length $\lambda_L=800$nm. The laser pulse impinges normally on a $40$nm -thick, sharp-edged fully ionized Carbon plasma foil initially located at $x=6\lambda_{L}$, with an initial electron density of $n_e=21n_c$. This configuration ensures a relativistic near-critical density regime, characterized by the similarity parameter $S=n_e/a_{0}^{d} n_{c}\approx1$ \cite{Gonoskov2011}, which is optimal for efficient electron blowout and RES formation. Here, the critical density is defined as $n_c=mc\omega^{2}_{L}/e^2 = 1.7 \times 10^{21}$cm$^{-3}$ where $m$ denote the electron mass. The initial temperature of the plasma electrons is set as $T_e=100$eV. The size of each spatial grid of the simulation box is $\Delta x=\lambda_{L}/4000$, in which there are $500$ macro-particles of each species.\\

As the sharp-fronted drive pulse reaches the target at $t=6T_{L}$, the radiation pressure at the leading edge of the pulse rapidly exceeds the electrostatic binding forces between electrons and ions. This results in a prompt and coherent expulsion of electrons in the forward direction, parallel to the laser propagation axis. The expelled electrons are compressed into a dense, ultra-thin RES that is accelerated synchronously as is evident in Figure~\ref{fig:gau_2w}(a). Although such an approach using non-adiabatic sharp rising pulse has been investigated in numerous occasions for generating RES, accessing such pulses is extremely complex and challenging. Considering the key role of pulse front steepness, plasma-based pulse front steepening was proposed in \cite{Wang2011} and implemented in ion acceleration experiments \cite{Bin2015}. In such experiments a separate low density target is designed prior to the main target, where relativistic non-linearities introduced by this near critical density plasma modifies the front edge of the pulse. This approach ultimately leads to a laser pulse, which is spatio-temporally compressed \cite{Bin2021}. Once such pulses are generated, a single RES can be formed \cite{Ma2014}. However, these methods, which are based on non-linear pulse shaping within a plasma medium, introduce a significantly higher level of experimental complexity. 

A typical currently available, ultrashort high power laser system delivers pulses with Gaussian temporal envelop on target. Figure \ref{fig:gau_2w}(b) presents the electron density evolution when the same plasma foil is irradiated by such a Gaussian-shaped laser pulse with identical peak field amplitude $a_0^{d}=20$. Despite having the same peak intensity, $I_{0}^{d}[Wcm^{-2}]=1.37\times 10^{18}(a_{0}^{d}/\lambda_{L}[\mu m])^2$, the gradual rise of the Gaussian envelope fails to deliver a sufficiently strong ponderomotive force at the pulse front. As a result, the electrons are not coherently swept into a single sheet but instead experience multiple oscillations within the laser field. This leads to incomplete electron blowout, in contrast to the case presented in Figure~\ref{fig:gau_2w}(a), and the plasma surface becomes distorted and modulated, rather than forming a clean RES. The electrons remain partially bound and oscillate around the ion background, resulting in a disordered plasma structure with reduced coherence and acceleration efficiency. It is also evident in Figure~\ref{fig:gau_2w}(b), that there are signatures of partial electron ejection on the back side of the irradiated target, leading to multiple potential electron bunches of varying strength. This comparison highlights the critical role of the pulse temporal profile—particularly the sharpness of the leading edge—in determining the efficiency of RES formation. A steep front ensures a rapid onset of radiation pressure, enabling synchronized electron expulsion, while a smooth Gaussian front leads to phase mixing and multiple bunch emission leading to their structural degradation.

This distinction in interaction dynamics is particularly relevant in the context of coherent Thomson backscattering for generation of \emph{as} pulses. In the experiment reported by Kiefer et. al. \cite{Kiefer2013}, a $10$nm diamond-like carbon (DLC) was irradiated by a Gaussian-shaped drive pulse with intensity exceeding  $1\times10^{20}$Wcm$^{-2}$. Due to the smooth rising edge of the Gaussian pulse, the initial few laser cycles were insufficiently intense to fully overcome the binding forces within the foil. Consequently, only a fraction of the electrons are expelled during each cycle of the laser field, resulting in the formation of multiple relativistic electron layers, or RESs, at the rear side of the target. These RESs, being only partially reflective, interact with a counter-propagating probe pulse of intensity $\sim 1\times10^{15}$Wcm$^{-2}$. As the probe pulse is back-reflected from a sequence of partially transmissive relativistic electron sheets, generating an attosecond pulse train. This process produces a reflected spectrum rich in high-order harmonics of the fundamental frequency of the probe. This mechanism stands in contrast to the interaction driven by a sharp-edged pulse front, which is capable of expelling all plasma electrons within a single half-cycle, thereby forming a single, highly reflective RES. This configuration is more favorable for generating isolated attosecond bursts.

Since the RES formation is directly governed by the drive pulse field, there exists a strong correlation between the waveform of the drive pulse and the coherent electron dynamics responsible for RES generation \cite{Kulagin2004, Kulagin2007,Kulagin2009}. Here, we demonstrate that the laser field waveform can be shaped optimally by the addition of a second harmonic component to a Gaussian drive pulse to excite plasma electron dynamics, that enables the formation and acceleration of a single, dense relativistic electron sheet. The significance of drive pulse waveform engineering has been widely explored in various high-field plasma interaction regimes, including laser wakefield acceleration \cite{Li2019} and attosecond XUV pulse generation via high harmonic emission from plasma surfaces \cite{Mirzanejad2013, Edwards2016,Yeung2016,Zhang2020,Chopineau2022,Mondal2022,Fitzpatrick2024}. These studies consistently highlight the importance of sub-cycle control over the laser field to manipulate electron trajectories and optimize emission characteristics.

In Figure~\ref{fig:gau_2w}(c), we present the case when the drive laser waveform is shaped by mixing with the fundamental frequency $\omega_{L}$, a fraction $P$ of it's second harmonic $2\omega_{L}$. As indicated by the arrow in Figure~\ref{fig:gau_2w}(c), this two-color linearly polarized drive configuration effectively excites plasma electrons which are compressed into a coherent structure forming a single RES. Figure~\ref{fig:gau_2w}(d) illustrates the two drive pulses employed to initiate interactions, presented in Figure~\ref{fig:gau_2w}(b-c), with different drive pulse waveforms: Gaussian (in black marked $1\omega$), and two-color Gaussian (in magenta marked $1\omega+2\omega$). However, the distinct temporal waveform of each pulse leads markedly different electron dynamics and RES formation behavior. For simulations presented in Figures~\ref{fig:gau_2w}(b) and \ref{fig:gau_2w}(c) total energy contained within the drive laser pulse is kept constant. 
\begin{figure*}[t!]
    \centering
    \includegraphics[width=1.75\columnwidth]{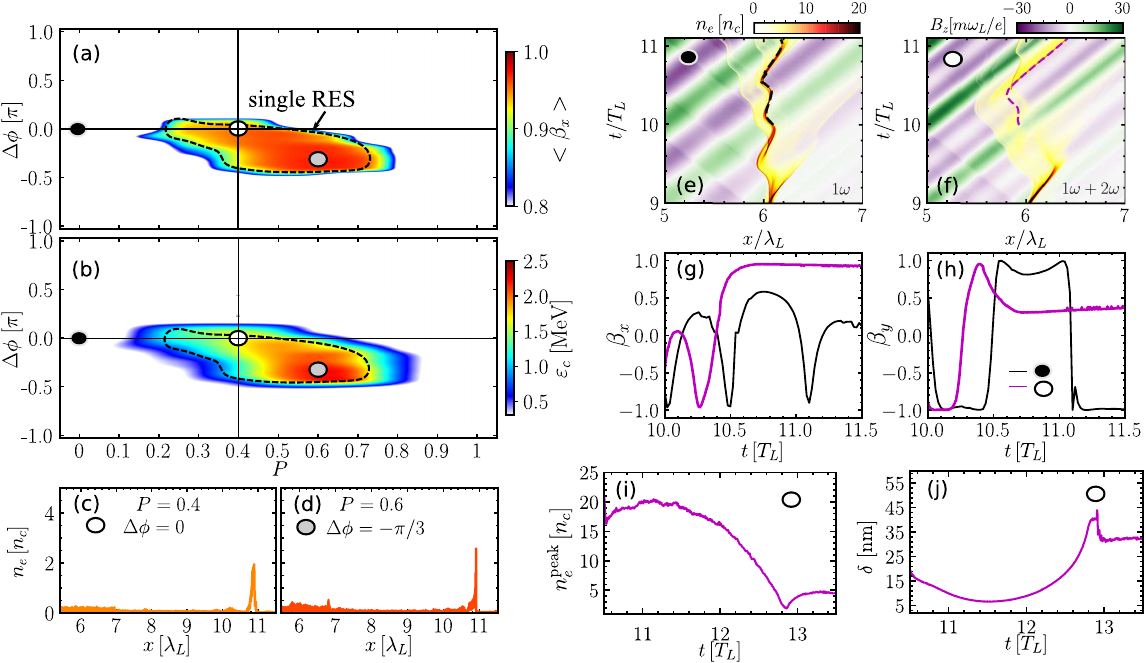}
    \caption{Optimal generation of single RES with waveform-controlled laser driver. (a) The longitudinal velocity of the ejected electron beam, averaged over the bunch $<\!\beta_{x}\!>$ as a function of $P$ and $\Delta \phi$. (b) The waveform dependent maximum energy of accelerated electrons in the bunch represented by the acquired cutoff energy $\varepsilon_{c}$ in ($P,\Delta \phi$)-space. The dashed encircled region in (a) and (b) denotes the parameter space, in which a single RES is produced and accelerated. The density profile of the electron beam after traveling $5\lambda_L$ to the location of detector $(x=11\lambda_{L})$ are plotted for two optimal waveforms of the drive laser (c) $P=0.4$ with $\Delta \phi=0$, and (d) $P=0.6$ with $\Delta \phi= -\pi/3$, in which a single electron beam with attosecond duration is emitted. (e) shows the spatio-temporal evolution of thin foil electron density (red to yellow colormap) and drive pulse field (green to violet colormap) for a single color Gaussian driver. (f) The spacetime evolution of the laser field and electron density for single RES regime as well as the temporal evolution of phasespace parameters of electron bunches. The time axis implies the simulation time (see Fig.~\ref{fig:gau_2w}(b) and (c)) during interaction of drive beam, synthesized by (e) only fundamental frequency ($1\omega$) or (f) by combination of in phase frequencies $(1\omega+2\omega)$ with energy ratio $P=0.4$. The dashed lines in (e) and (f) represent trajectory of the representative electrons which are subject to the normalized drive field $B_z$. (g) and (h) show temporal variation of the longitudinal and transverse components of the velocity of the representative electrons (black curve for $1\omega$ and magenta for $1\omega+2\omega$ field). The temporal evolution of (i) bunch peak density and (j) thickness during the interaction with a counter propagating source pulse with the RES generated by the two-color Gaussian driver beam (with $P=0.4$ and $\Delta \phi= 0$). The FWHM of the bunch is calculated by fitting a Gaussian function on the electron density profile. Greater longitudinal velocity of RES during interaction phase accounts for larger compression of the AP in case of synthesized pulse.}
    \label{fig:e_charge}
\end{figure*}

Once a single forward propagating RES is formed, a cross-polarized, counter-propagating source pulse is introduced at a carefully timed delay relative to the drive pulse. This source pulse interacts with the RES at normal incidence and is reflected, undergoing a Doppler frequency upshift and temporal compression into a single attosecond pulse (AP). The efficacy of RES for providing optimal conditions for TBS of a linearly polarized Gaussian source pulse (electric field along the  z-axis), with normalized amplitude with $a_0=0.1$ and a temporal duration of $4$fs (corresponding to $1.5T_L$ in FWHM of intensity) is summarized in Figure~\ref{fig:gau_2w}(e-f). This correlation between the quality of RES and back reflected pulse, is quantitatively illustrated in Figure~\ref{fig:gau_2w}(e), which compares the temporal profiles of TBS pulses generated by RESs formed under different drive pulse temporal waveforms. The two-color drive pulse produces a compressed AP with a duration of (intensity FWHM) $110$ \emph{as} and a peak intensity of $\approx 2.6 \times 10^{15}~Wcm^{-2}$ (corresponding to $a_0^{2} \approx 1.2 \times 10^{-3}$), significantly outperforming the weaker and longer AP ($380$as) generated by even the ideal top-hat pulse. This clearly demonstrates the enhanced efficiency and coherence enabled by waveform shaping. The temporal profile for the Gaussian case does not lead to single AP formation.

The corresponding frequency spectra presented in Figure~\ref{fig:gau_2w}(f) further reinforce this conclusion. The ideally top-hat drive pulse is plotted in blue curve and bicolor Gaussian pulse in magenta curve. The RES formed by the two-color drive pulse reaches a higher velocity, resulting in a larger central frequency shift of the reflected spectrum up to $26\omega_L$ and a broader bandwidth compared to the sharp rising pulse front case, ideal for shorter AP. In contrast, in case of Gaussian drive (black curve in Figure~\ref{fig:gau_2w}(f)), most of the scattered energy is concentrated in low order harmonics induced in the reflected pulse, in line with the observation discussed in Figure~\ref{fig:gau_2w}(b).

These results underscore the critical role of multi-frequency waveform synthesis in controlling RES formation and optimizing attosecond pulse generation. A deeper analysis of this mechanism is presented in Section~\ref{sec:singleREM}, where the dynamics of single RES formation and its impact on coherent backscattering are discussed in detail.  Once single RES is formed, it can be combined with double-layer target configuration \cite{Wu2010a,Wu2012a,Ma2024}, to boost the longitudinal Lorentz factor $\gamma_{x}=(1-\beta_{x}^{2})^{-1/2}$ further and enhance the backreflected \emph{as} pulse, where $\beta_{x}$ is the longitudinal component of the RES velocity in units of $c$.

\section{Generation of Single RES by controlling driver waveform}\label{sec:singleREM}

In order to investigate light-wave controlled single RES regime, and its application for subsequent single AP generation, here we undertake a multiparametric study.
A Gaussian temporal profile of $g(\zeta)=\exp(-2\ln2(\zeta-6)^2/\tau_P^2)$ with FWHM duration of $\tau_p=3.7T_L=10$fs is considered, where $\zeta=t-x/c$, being the retarded coordinate. The linearly polarized driver field, $E_y^{d}=E_y^{1\omega}+E_y^{2\omega}$, is synthesized from superposition of the fundamental frequency $1\omega=\omega_L$ and its second harmonic $(2\omega)$. The normalized electric field of $1\omega$ beam is, $eE_{y}^{1\omega}(\zeta)/mc\omega_L=a_{0}^{d}\sqrt{1-P} g(\zeta)\cos(\omega_L\zeta)$.
The second harmonic component $2\omega$  is presented by,
$eE_{y}^{2\omega}(\zeta)/mc\omega_L=a_0^{d}\sqrt{P} g(\zeta) \cos(2\omega_L\zeta+\Delta\phi)$.
Here the parameter $P$ denotes the portion of the total energy, which is converted to its second harmonic beam. For any energy contribution of the second harmonic component to the total pulse energy $P$ and $a_{0}^{d}=20$, the total pulse energy (power averaged over time) of drive beam is preserved in each case. $\Delta \phi$ stands for the relative phase between the fundamental frequency $1\omega$ and its second harmonic $2\omega$ component. We have shown previously that RES formation is extremely sensitive to laser waveform shaping. Thus, a finely adjusted drive pulse electric field might enable us to control the ejection and acceleration of plasma electrons. 

Here we undertake a detailed study of the influence of light-wave control on RES formation by fine-tuning the electric field of the drive laser with the experimentally accessible parameters ($P,\Delta\phi$). Figure~\ref{fig:e_charge}(a-b) summarizes the outcomes of 407 simulations in the ($P,\Delta\phi$)-space, where each run corresponds to one synthesized driver waveform spanning from single color $1\omega$ case through double color ($1\omega+2\omega$) to single color $2\omega$. The map on phase difference ranges from $-\pi$ to $\pi$, which is the periodicity of the total drive field.

In order to grasp a meaningful understanding, first we identified and separated the domain of single RES formation from that corresponding to generation of zero or multiple RES. The region marked with the black dashed line in Figure~\ref{fig:e_charge}(a-b) highlights the generation of single and overdense RES under the action of driving beam. The waveform representing the Gaussian driver corresponding to the case presented in Figure~\ref{fig:gau_2w}(e) is marked in black filled circle with gray contour, in Figure~\ref{fig:e_charge}(a-b), which in this case lies outside the black dashed area. The two-color Gaussian pulse with various energy ratio of $P$ and $\Delta\phi$, shows a contiguous region of waveforms where the drive laser interaction yields high energy electron jets  emitted toward the rear side of the target. 

Figure~\ref{fig:e_charge}(a) plots the average $\beta_{x}$ (longitudinal speed normalized to $c$) of the electrons within bunch moving with relativistic longitudinal velocity, $<\beta_{x}> = \int_{0.8}^{1} (\frac{dN}{d\beta_{x}})\beta_{x} d\beta_{x} / \int (\frac{dN}{d\beta_{x}}) d\beta_{x}$, analyzed at location $x=11\lambda_{L}$, about $5\lambda_{L}$ away from target rear surface.  Since for stable \emph{as} pulse generation we need a RES regime where the generated electron sheet sustains for longer duration, for this analysis we consider the properties of the RES approximately 10$T_{L}$ after the interaction is initiated (roughly the time the RES generated in Figure~\ref{fig:gau_2w}(c) takes to reach $x=11\lambda$). The cut off energy $\varepsilon_{c}$ associated with the corresponding energy map is presented in Figure~\ref{fig:e_charge}(b) as a function of synthesized driving field features $P$ and $\Delta \phi$. The results show that single sheet of very high density can be accelerated up to $2.5$ MeV cutoff energy through light-wave shaping in this parameter range. In Figure~\ref{fig:e_charge}(c-d) we plot two typical spatial distributions of the electron density in the single RES regime. The RESs generated by the Gaussian drive field with energy ratio around $P=0.4$ and in phase frequency components, i.e. $\Delta \phi=0$ (white circle with black contour) and $P=0.6 $ with phase difference of $-\pi/3$ (gray circle with black contour) are shown. The corresponding waveforms are marked on Figure~\ref{fig:e_charge}(a-b) colormaps. The spatial distribution of the electron density demonstrates the generation of single RES in the target rear side. 

At a constant $P$, as the phase difference between two frequency components of the synthesized driving field varies from zero, the steepness of the waveform will change. While converting more energy to the second harmonic will modify the temporal width of the modulated part of the drive field once the phase difference is kept the same. Thus, the reason of obtaining high energy and dense electron beam at this parameters can be attributed to the shape of the laser waveform interacting with the target. Adding a particular portion of second harmonic to the fundamental beam modulates the waveform of the laser field, ensuring proper coupling and sufficient push force that leads to blow electrons out from the plasma foil under a specific condition. 

As can be seen from two-color field waveform at the instant of electron acceleration (marked by an arrow in Figure~\ref{fig:gau_2w}(d)), the steeper slope of the drive field exerted on the electrons gathers electrons into a single bunch and pushes them faster (higher acceleration gain) compared to $1\omega$ field to reach relativistic velocities as presented in Figure~\ref{fig:e_charge}. Figure~\ref{fig:e_charge}(e) and (f) depict the spatio-temporal evolution of electron density of the foil  (yellow to red color bar) overlaid with the magnetic field component of the drive pulse (green to violet color bar) during interaction. The representative electron trajectories under the influence of these fields are indicated as dashed lines in Figure~\ref{fig:e_charge}(e-f).

Two distinct sub-cycle electron dynamics emerge in the two regimes, subject to different drive fields and electrostatic field evolution. Electrons perform oscillation around ion profile in case of single color field as can be seen in Figure~\ref{fig:e_charge}(e). Partial RES emission for the single colour Gaussian pulse, can be seen at 9$T_{L}$ in Figure~\ref{fig:e_charge}(e), along with oscillatory electron motion around the background ion core. In contrast, in the case of an optimal two-color field, the electrons are blown out (Figure~\ref{fig:e_charge}(f)) and then gain forward acceleration. As illustrated in Figure~\ref{fig:e_charge}(f), in the first few cycles at the front edge of the synthesized field, since the ponderomotive push force is not sufficient to overcome the electrostatic force, the blown electrons, which are compressed until $9.6T_L$, will decompress and therefore be pulled back to the front side of the target ($x<6 \lambda_L$), in which both the ponderomotive force and the electrostatic force push and compress the electrons into a single RES. Such an electron bunch with initial relativistic velocity induced by electrostatic force of ions is able to interact directly with the drive pulse field and can acquire further longitudinal energy. These re-bunched electrons around simulation time $10.4T_L$ are subject to a persistent push force, which is unipolar and does not change its sign, therefore get accelerated to relativistic velocities (magenta line in Figure~\ref{fig:e_charge}(g)), until the laser overtakes the electron bunch. Eventually, the RES, which is phase-locked with the drive field and is accelerated until dephasing occurs, will fly for a prolonged time and create a robust reflective electron layer with relativistic velocity. The synergistic correlation between driving laser waveform and plasma electrons gives rise to such electron bunch dynamics and it is typical of the single RES regime presented in Figure~\ref{fig:e_charge}(a-b).

The role played by the longitudinal restoring force due to the space charge separation is crucial for dual-color drive beam particularly when the plasma electrons are expanded to the left hand side of the plasma ion slab, then being pushed to the forward direction, such that electrons within the left end of the bunch undergo larger amount of push force (electrostatic plus light pressure) and therefore can catch the slower electrons in the right side of the bunch. The electrostatic force accelerates them to relativistic velocity, the moment after which the RES is injected into the drive pulse field and experiences an unipolar field. The trapped bunch in the transmitted laser field is able to gain energy until it dephases \cite{Thvenet2015,Zhou2020}. 

To understand the mechanism better, in Figures~\ref{fig:e_charge}(g) and \ref{fig:e_charge}(h), we compare the temporal evolution of longitudinal and transverse components of velocity of the electrons involved in the interaction with single (black line) and two-color (magenta line) drive pulses from simulation time $10$ to $11.5T_L$, when the final stage of RES formation and acceleration occur. In case of $1\omega$ laser beam the electrons longitudinal velocity $\beta_x$ (black line in Figure~\ref{fig:e_charge}(g)) reaches a peak in positive direction and then drops, every half cycle, since the laser push force is not sufficient for complete blowout.
In Figure~\ref{fig:e_charge}(h) the transverse component $\beta_{y}$ (magenta line), of the electron velocity, under the mpact of two-color field peaks at $t=10.4T_{L}$, where the persistent laser magnetic field experienced by the electrons within the bunch exerts a longitudinal force via $\sim \beta_{y} B_{z}\hat{x}$. In case of two color drive pulse, both phases of expelling and accelerating of electrons take place within nearly half laser cycle, during which field does not change sign from $t=10.1T_{L}$ to $11.7T_{L}$. This field waveform with steeper slope not only avoids any additional electron oscillation, but also imposes an increasingly push force on bunched electrons, leading to synchronously reach relativistic velocity along positive x-axis (magenta line in Figure~\ref{fig:e_charge}(g)). As the result of this larger longitudinal velocity of the RES and its lower energy spread, the temporal duration of the scattered source pulse is reduced by $3$ fold.
\begin{figure}[!t]
    \centering
    \includegraphics[width=1\columnwidth]{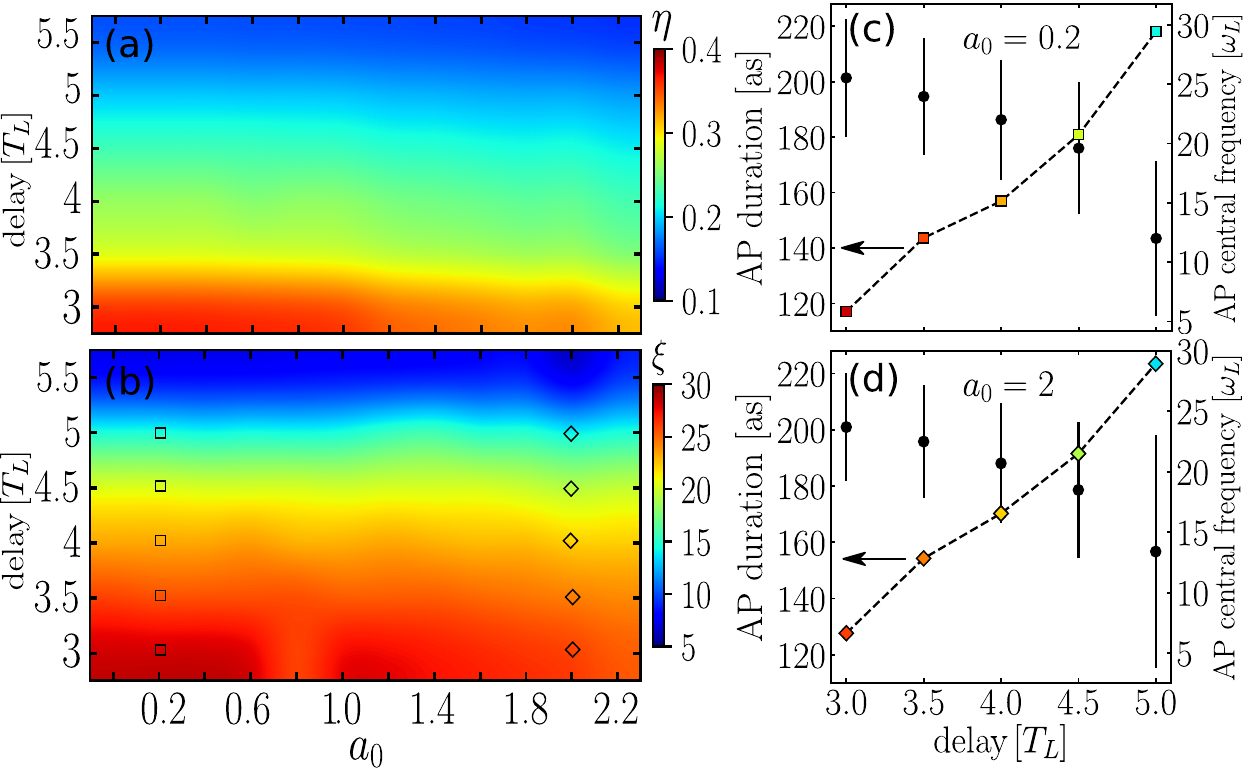}
    \caption{The attosecond back-scattered pulse generated by in phase two-color drive beam with energy mixture ratio of $P=0.4$. The delay-amplitude maps for a $4$fs Gaussian source beam at a given amplitude $a_0$. (a) and (b) present scaling for Thomson back-scattered (TBS) pulse peak amplitude ratios $\eta=a_0^{TBS}/a_{0}$ and temporal compression $\xi=\tau/\tau_{_{TBS}}$ compared to the counter-propagating source pulse with amplitude $a_0$ and duration $\tau=4$fs. Temporal duration of each Thomson back-scattered field overlaid with its central frequency for the case of source pulses $a_{0}=0.2$ (c) and $a_0=2$ (b) are depicted. The TBS spectra consist of a peak (central) frequency with a FWHM bandwidth, indicated by a vertical bar. In each case, the FWHM is calculated by fitting a Gaussian function on the squared field and spectrum.}
    \label{fig:delay_maps}
\end{figure}
Signatures of attosecond electron bunch acceleration from plasma mirror targets have been captured in recent experiments and it has been shown that the process sensitively depends on the proper injection of the electron bunch in proper phase of the copropagating drive laser field \cite{Thvenet2015,Zhou2020}. Due to the distinct feature of drive pulse structure, the injected electrons are required to acquire a proper initial velocity to be trapped in the electromagnetic field of the drive laser fields without changing sign. The electrons are accelerated to the relativistic velocity ($0.8c$) and are then injected into the drive laser field so that they form a nm scale bunch, which undergoes persistent laser field and advancing in phase for a long period of time. This trapped relativistic electron bunch in the laser field gains a significant amount of energy, giving rise to approach longitudinal velocity close to $0.95c$. Eventually, the laser push force takes over the acceleration process of electrons, and so the acceleration force maintains for a longer time.

In order to gain more insights into the evolution of RES properties we plot in Figure~\ref{fig:e_charge}(i-j) electron bunch peak density and RES width for two-color Gaussian driver for the case of ($P=0.4,\Delta \phi=0$). The peak electron density can go upto $20n_{c}$ as seen in Figure~\ref{fig:e_charge}(i). Due to the creation of highly dense RES, the intensity of the reflected source could be enhanced by more than one order of magnitude, as we have already presented in Figure~\ref{fig:gau_2w}(e). Figure~\ref{fig:e_charge}(j) shows that the generated RES is sub-$10$ nm in thickness and implies that the bunch preserve its thickness during interaction with the source pulse. For the case of two-color beam, at $\sim12.8T_L$ the intensity of the driving pulse drops which causes the electron bunch repulsion force to grow, therefore the RES density decreases to around $2n_c$. However, then by increasing the drive field value, the RES is again piled up.  

\section{Control of attosecond pulse properties in single RES regime}

In the single RES regime achievable through light-wave control and investigated in the previous section, we notice that representative features that characterize the accelerated electron sheet (thickness, mean energy and peak electron density etc) evolve with time. This implies that two crucial aspects that are of utmost importance for isolated \emph{as} pulse generation through TBS, are the instant at which the back-propagating ultrashort pulse starts interacting with the RES and the evolution of the RES within the interaction window. Thus, the delay between interaction of drive and source beams with the foil is of prime importance, and affects on the features of the scattered radiation. Another parameter that is of importance is the peak electric field of the incident source pulse represented by normalized amplitude $a_{0}$. Hence, we utilize this delay and $a_{0}$ as two control parameters for optimizing the properties of the generated attosecond pulse. As presented in Figure~\ref{fig:e_charge}, a dual-color drive pulse with energy ratio parameter of $P=0.4$ and $\Delta \phi =0$ is able to eject plasma electrons coherently to transmitted direction of the foil, producing a RES suitable for Thomson scattering of a counter-streaming source pulse. Without loss of generality, we investigate the AP characteristics generated from this RES, as a scenario representative of the single RES regime presented in Figure~\ref{fig:e_charge}(a-b).

The accelerated RES is irradiated by a counter propagating source pulse, having a crossed linear polarization along z direction $(E_z)$. The backscattered field is an isolated attosecond pulse in of 110$as$ duration as shown in Figure ~\ref{fig:gau_2w}(e). For representing the AP properties we have defined two parameters. The generated \emph{as} pulse field enhancement is captured through peak amplitude ratios $\eta=a_{0}^{TBS}/a_{0}$, where $a_{0}$ is the normalized field amplitude of the incident source pulse and $a_{0}^{TBS}$ is the same for the \emph{as} pulse. We also define temporal compression $\xi=\tau/\tau_{_{TBS}}$, where $\tau = 4$fs is the duration of the  Gaussian source pulse and $\tau_{_{TBS}}$ is the pulse duration of the generated \emph{as} pulses. 

Figure~\ref{fig:delay_maps} shows the parametric scan of the features of reflected attosecond burst with respect to delay and amplitude of the Gaussian $\tau = 4$fs source pulse. Here, the zero delay denotes the time, at which both the drive and source pulses arrive at the position of foil target at the same time ($t=6T_{L}$ in simulation time). In Figure~\ref{fig:delay_maps}(a) the ratio of peak amplitude of scattered AP to that of incident source $\eta$ is presented in colormap as a function of delay and $a_0$. Panel (b) in Figure~\ref{fig:delay_maps} illustrates the achieved temporal compression value $\xi$ of the scattered pulse with respect to the $\tau = 4$fs source pulse with variation in delay at different $a_0$. We observe that both $\eta$ and $\xi$ has strong dependence on delay and varies weakly with $a_{0}$. The highest compression ratio and field enhancement can be achieved at shorter delays for a constant $a_{0}$. Figure~\ref{fig:delay_maps}(a) shows that for $a_{0}<1$ (sub-relativistic intensities) $\eta$ shows a similar dependence on delay. As the source pulse strength is increased beyond the relativistic limit $a_{0}>1$ the enhancement factor $\eta$ reduces at the same delay, i.e for the same RES properties, indicating that the RES starts to be perturbed at such source pulse intensities. To get a deeper insight, we look into the AP properties as a function of delay at a non-relativistic ($a_{0}=0.2$) and relativistic ($a_{0}=2$) source pulse intensities.

The variation of the temporal durations of scattered AP for source $a_0=0.2$ and $a_0=2$ are shown respectively in Figure~\ref{fig:delay_maps} (c) and (d) (the curves with arrow pointing to the left axes). The spectral domain features with the black dots representing the central frequencies and the bars indicating the corresponding spectral bandwidths are also presented in  Figure~\ref{fig:delay_maps} (c-d) with values on the right axes. The characteristics of the reflected source is strongly dependent on the delay time. Therefore, by imposing a proper delay one can adjust central frequency of the scattered attosecond pulse. A delay time of $3T_L$ favors the generation of single AP with shorter temporal duration and narrower bandwidth at higher frequencies. Up to certain limit by increasing the intensity of the source pulse the intensity of the scattered signal will accordingly enhance. In case of relativistically intense source pulse $a_{0}=2$ at any delay of Fig.~\ref{fig:delay_maps}, the temporal duration of APs become larger, which can be explained by less longitudinal velocity of the RES due to opposite push force of the source beam. 

\begin{figure*}[tb]
    \centering
    \includegraphics[width=1.85\columnwidth]{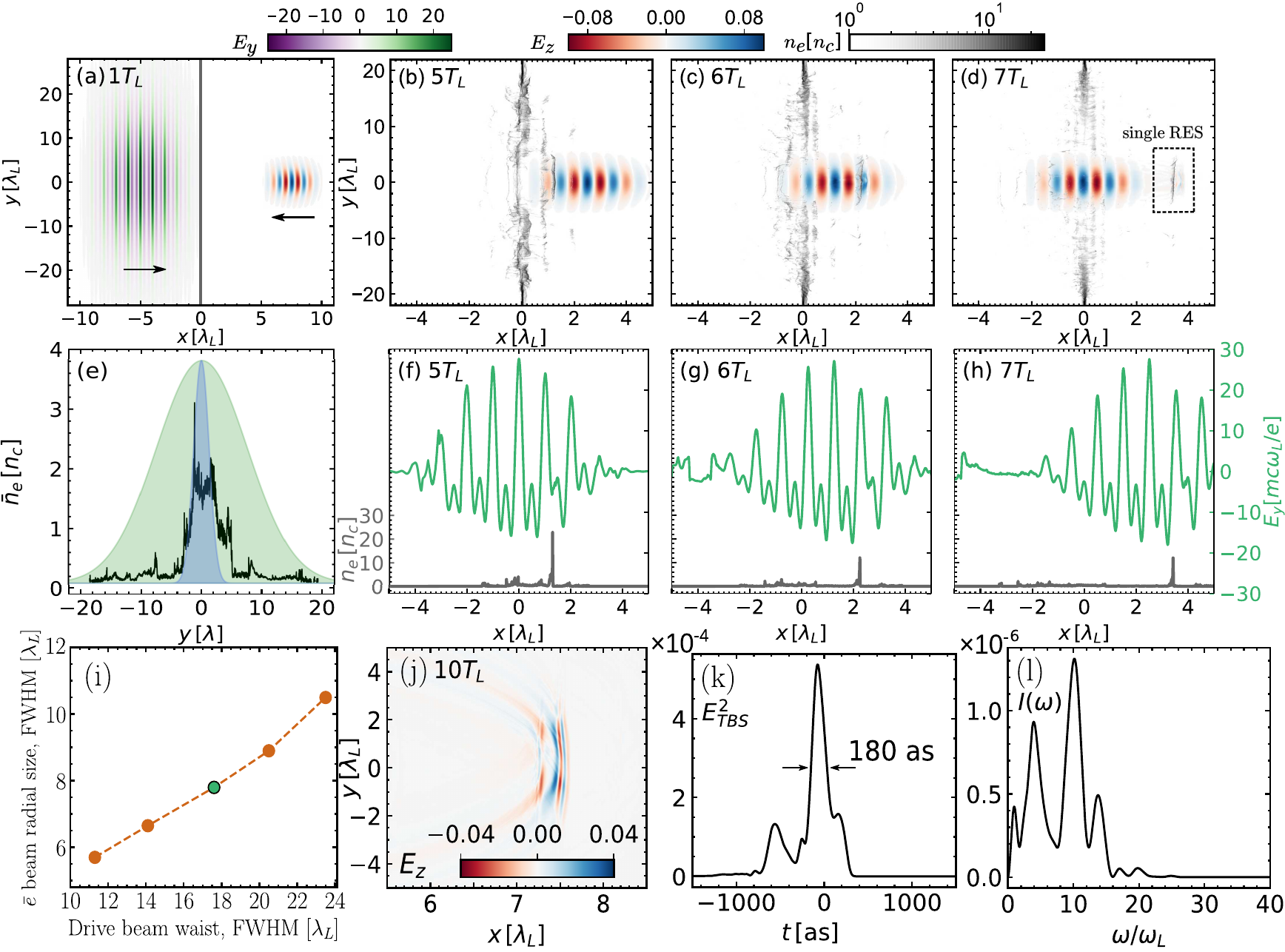}
    \caption{Geometry of the dual-color Gaussian drive beam (represented by electric field $Ey$ proceeding to the right) and counter-propagating source beam (with electric field $E_z$ propagating to the left) just before the interaction at $1T_{L}$ in (a). Electric field magnitudes are normalized to $mc\omega_L/e$. The dynamics of the interaction is shown (b)-(d) where the single RES is accelerated in the forward direction and reflects a fraction of the source pulse marked by a rectangular box in (d). Panels (f)-(h) depict the on-axis spatial lineout of the drive pulse (green lines) overlaid with the electron density (gray lines) at $y=0$ averaged over $7$ adjacent grid cells. The distribution of the the single RES in transverse direction at $5T_L$ (black line) is demonstrated in (e) overlaid with the normalized transverse profile of the source pulse (light blue shade) and the initial transverse profile of the drive pulse on target in light green shade. Spatial, temporal and spectral features of the back-scattered source pulse reflected by the single RES generated by two-color Gaussian drive beam at simulation time $10T_L$ are depicted in panels (j)-(l), respectively. The temporal profile of the squared field is shown in (k) with $180$\emph{as} duration at its FWHM, and the corresponding modulated spectrum in (l) without performing any frequency filtering. Maintaining all the interaction parameters, the radial size dependence of the generated single RES on the focal spot of the drive laser is plotted in panel (i). This analysis is performed at the moment $5T_{L}$ when the source pulse is supposed to interact with the RES. The green color coded dot represents the case in panel (e).}
    \label{fig:2d_geo}
\end{figure*}

\section{Formation of single RES and AP generation with finite focal spot size}

In order to pinpoint not only the temporal and spectral characteristics of the scattered pulse but also its spatial features we have performed simulations, incorporating finite  and realistic laser focal spot size , with PIC code, PICCANTE \cite{Sgattoni2015}. The simulation box which is from $-30\lambda_L$ to $30\lambda_L$ (in both $x$ and $y$ direction) consists of $60000\times 12000$ cells, in laser propagation direction and its normal, respectively. In each cell we have employed $60$ macro-electrons and $10$ macro-ions of fully ionized Carbon. The linearly polarized drive beam with electric field $E_y$ and central wavelength $\lambda_L=800$ propagates from left to right and impinges the pre-ionized $40$nm plasma foil located at $x=0$. Considering transverse Gaussian profile of the intensity $\propto \exp(-4\ln{2}y^{2}/w_0^{2})$, the beam spot size at target position is $w_{0}=18$\textmu m (in FWHM of intensity), as well as a Gaussian temporal profile of duration  $10$fs at FWHM of intensity envelope. This multi-cycle drive beam with normalized field strength of $a_{0}^{d}=20$ is temporally and spatially combined with its second harmonic with a ratio of $P=0.4$ which resembles the optimal drive waveform obtained from 2D plane wave simulations presented previously.

Figure \ref{fig:2d_geo}(a) depicts the configuration of the interaction where the intense dual-color drive beam reaches the foil position at $1T_L$ simulation time. The initial electron density of the plasma slab is set to be $32n_{c}$, which is slightly greater than of that in $2$D plane wave simulations ($21n_{c}$). This is due to the fact that the relativistic ponderomotive force associated with the focused drive pulse introduces a transverse scattering push on the plasma electrons and expels them to the region, at which light intensity is lower. Therefore, the effective electron density of the plasma, and consequently the electrostatic field at the interaction surface will be reduced. This effect is thus compensated by increasing the initial electron density of the target to maintain the proper interaction conditions for the generation of single RES. The focused drive beam expels the plasma electrons and accelerates them in phase along its propagation direction. After a delay of $3T_L$, since when the leading edge of the drive has reached the foil front position, a low intensity counter-propagating source pulse with cross-polarized electric field $E_z$ interacts with the driven RES, which travels in positive $x$ direction. The tightly focused Gaussian source beam has $4$fs temporal duration and $2.9\lambda_{L}$ transverse dimension in FWHM of its envelope on target at $x=0$. At this instant of interaction, the plasma slab which is subject to intense electromagnetic field of the drive beam turns relativistically transparent, leading to the transmission of drive pulse. The temporal evolution of the single RES generation process is shown in Figure~\ref{fig:2d_geo}(b)-(d), highlighting the reflection of a portion of the source pulse by the RES. The transverse extent of the generated single RES is illustrated in Figure.~\ref{fig:2d_geo}(d) in black line. Figure.~\ref{fig:2d_geo}(d) shows that the source pulse is covered by the size of the RES formed. The lineout at $y=0$ (averaged over $0.2 \lambda_{L}$) of the drive beam field $E_{y}$, which is plotted in Figure~\ref{fig:2d_geo}(f) to \ref{fig:2d_geo}(h) in green curve, substantiate the fact that analogous to previous results, plasma electrons are ejected forward by the same cycle of the drive pulse. Afterward, the accelerated RES and drive pulse propagate together in forward direction. Since, the drive pulse is loosely focused, the co-moving RES undergoes negligible amount of ponderomotive scattering force, maintaining its structure for a longer time.

The produced RES retains its radial structure during propagation for a few wavelengths, which is suitable for reflecting the source pulse. The influence of focal spot size of the drive laser on the spatial extent of the generated single RES is illustrated in Figure~\ref{fig:2d_geo}(i). This provides insight for adjusting the appropriate focusing spot size of the counter-propagating source beam. Other simulation parameters are kept the same. The transverse size of the RES is assessed at $t=5T_{L}$, when the source pulse begins interacting with the RES. Figure~\ref{fig:2d_geo}(i) also demonstrate that the proposed light-wave control for single RES formation is a robust mechanism that stays insensitive to drive focal spot size variation.

As a consequence of interaction of this RES and counter-propagating source beam a single sub fs radiation, consisting of high frequencies is generated and propagates in forward direction with polarization similar to that of source beam. The scattered portion of the source pulse, which advances to the positive $x$-axis is illustrated in Figure~\ref{fig:2d_geo}(j-l). The spatial dimension of this attosecond burst in Figure~\ref{fig:2d_geo}(j) also corroborates the generation of a single RES during interaction of a dual-color Gaussian drive pulse. Panel \ref{fig:2d_geo}(k) plots the envelope of the squared field with temporal duration of $180$as, which is qualitatively consistent with the previous results. The structured spectrum corresponding to lineout of $E_{z}$ (averaged over $0.1\lambda_{L}$) is shown in Figure~\ref{fig:2d_geo}(l), comprising a central peak at $10\omega_{L}$ and a low energy portion extending up to $20\omega_{L}$. 

\section{Summary and Conclusion}\label{sec:conclusion}

This study establishes a robust and tunable scheme for generating isolated attosecond pulses in the XUV regime via coherent Thomson backscattering from a single relativistic electron sheet. The formation of such a RES—a prerequisite for efficient frequency up-conversion—has long posed a challenge, particularly under conventional single-color Gaussian drive pulses that tend to produce multiple or poorly defined electron sheets. 

Through comprehensive particle-in-cell simulations, we demonstrate that a carefully synthesized two-color drive pulse, combining the fundamental frequency with its second harmonic, enables the controlled generation of a single, dense RES. This RES acts as a relativistic mirror, coherently reflecting and compressing a counter-propagating source pulse into a bright, attosecond burst. The temporal shaping of the drive field is shown to be critical, governing both the formation and acceleration dynamics of the RES. Moreover, we show that by adjusting the temporal delay between the RES and the source pulse, the carrier frequency of the scattered AP can be precisely tuned. 

This tunability, combined with the demonstrated robustness of the scheme across multiple simulation geometries, underscores its experimental viability. These findings mark a significant step forward in the field of relativistic optics, highlighting the power of laser waveform synthesis in controlling ultrafast plasma dynamics. The ability to generate bright, isolated attosecond pulses with high spectral purity and tailored properties promises new avenues for attosecond spectroscopy, ultrafast solid-state physics, and high-field science. The results are also important for high density electron bunch acceleration in the few MeV regime, which have important radiobiological applications \cite{RoadmapRad2025}. These developments also underscore the importance of synergy between advanced target fabrication and laser waveform engineering, for enabling a new regime of precision relativistic optics at the nanoscale.

The proposed scheme is grounded in two key recent advancements. 
First, the laser technology required to experimentally realize this approach is rapidly advancing \cite{Li2022,Han2023,Kubullek2025}, with high-intensity, waveform-controlled pulses becoming increasingly available at facilities such as ELI \cite{Chalus2024,Charalambidis2017,Shirozhan2024,Kahaly2025} and other laboratories worldwide \cite{ChLas2025,Veisz2025}. Second, innovation in ultrathin target fabrication has enabled relativistic laser-plasma interactions using few-nanometer-thick large-area carbon-based targets \cite{Kuramitsu2022} or replenishable microfluidic systems \cite{Koralek2018,Kim2023,Streeter2025}. These targets offer a unique combination of ultrathin geometry, high repetition rate compatibility, and surface smoothness, making them ideal for precision-controlled interactions with intense laser fields. When integrated with state-of-the-art, waveform-controlled laser systems, these ultrathin targets offer a unique testbed for investigating light-wave-driven electron dynamics—a central focus of this study. The ability to tailor the temporal and spatial structure of the laser field at sub-cycle resolution opens new avenues for coherent control of plasma surfaces, efficient RES formation, and attosecond-scale electron bunch generation.

\clearpage
\section*{Acknowledgements}
The ELI ALPS project (GINOP-2.3.6-15-2015-00001) was supported by the European Union and co-financed by the European Regional Development Fund. S.K. and M.S. acknowledge the High Performance Computation (HPC) facility at ELI-ALPS.
\section*{References}
\bibliography{REMref.bib} 

\end{document}